\documentclass[
]{ceurart}

\usepackage{natbib}

\usepackage{listings}
\usepackage{todonotes}
\usepackage{markdown}

\begin{document}

\copyrightyear{2025}
\copyrightclause{Copyright for this paper by its authors.
  Use permitted under Creative Commons License Attribution 4.0
  International (CC BY 4.0).}

\conference{Beyond Algorithms: Reclaiming the Interdisciplinary Roots of Recommender Systems Workshop (BEYOND 2025), September 26th, 2025, co-located with the 19th ACM Recommender Systems Conference, Prague, Czech Republic.}

\title{What News Recommendation Research Did (But Mostly Didn't) Teach Us About Building A News Recommender}

\author[1]{Karl Higley}[%
orcid=0009-0002-6332-8997,
email=khigley@umn.edu,
]
\address[1]{Department of Computer Science and Engineering, University of Minnesota}

\author[2]{Robin Burke}[%
orcid=0000-0001-5766-6434,
email=robin.burke@colorado.edu,
]
\address[2]{Department of Information Science, University of Colorado, Boulder}

\author[3]{Michael D. Ekstrand}[
orcid=0000-0003-2467-0108,
email=mdekstrand@drexel.edu,
]
\address[3]{Department of Information Science, Drexel University}

\author[4]{Bart P. Knijnenburg}[
orcid=0000-0003-1341-0669,
email=bartk@clemson.edu,
]
\address[4]{School of Computing, Clemson University}

\begin{abstract}
One of the goals of recommender systems research is to provide insights and methods that can be used by practitioners to build real-world systems that deliver high-quality recommendations to actual people grounded in their genuine interests and needs. 
We report on our experience trying to apply the news recommendation literature to build POPROX, a live platform for news recommendation research, and reflect on the extent to which the current state of research supports system-building efforts.
Our experience highlights several unexpected challenges encountered in building personalization features that are commonly found in products from news aggregators and publishers, and shows how those difficulties are connected to surprising gaps in the literature. Finally, we offer a set of lessons learned from building a live system with a persistent user base and highlight opportunities to make future news recommendation research more applicable and impactful in practice.
\end{abstract}


\maketitle

\section{Introduction}

\begin{quote}
    \textit{Everyone has a plan until they try to build a real system.} --- adapted from Mike Tyson \citep{warnerTallManTyrell1987}.
\end{quote} 

Recommender systems is a strongly applied research field, which draws people from many disciplines (including machine learning, human-computer interaction, information retrieval, psychology, marketing, economics, and more) interested in a shared class of problems: helping people discover information, products, and other items that meet their needs in a personalized way.
The community prides itself that the flagship RecSys conference attracts hundreds of industry researchers and practitioners each year, and encourages paper authors to pay attention to practical details like scalability, performance, and data quality.\footnote{These aspects have historically been emphasized by the Call For Contributions.} In practice, however, recommender systems research often does not provide the full set of tools needed to build effective recommender systems applications, as has been noted previously \citep{beel2017realworld, higley2022merlin}.
At the same time, reports from industry indicate that typical real-world recommender system deployments are quite simple and adopt few of the advanced techniques explored in the research literature \citep{shah2024adoption}.

Over the past two years, we have built a small-scale production recommender system, attempting as best we could to apply current best practices from the literature. We found gaps that go beyond the oft-discussed aspects of data management, deployment, and user experience. Many published results examine only one component of recommender systems (typically models), or single phases of the lifecycle of a user-recommender relationship (often after users have established interaction histories.) Published models are surprisingly difficult to apply to the kinds of data found in real-world datasets and practical recommendation problems. Many approaches for addressing issues like bias, fairness, and diversity conflict with each other by intervening through the same system components (e.g. re-rankers) or pursuing conflicting goals, so they can not readily be used in conjunction with each other. In sum, it proves difficult to assemble a working system from the components that have received research attention. The result is that the substantial and growing body of recommender systems literature, taken as a whole, is likely having less impact on improving real people's experiences with the recommendations they receive in their daily lives than the field would like.

In this case study, we explore these disparities between research and practice, grounding the findings in our effort to build a news recommendation research platform. We briefly introduce the project and its goals, along with a short survey of existing news recommendation research. We then describe several specific challenges we faced (and still face) while creating and operating it --- challenges that we expected the research literature to provide insights and techniques for, but found that it did not. We conclude with lessons and recommendations for fostering and producing research that is better able to deliver positive recommendation experiences for the real users of production recommender systems.

\section{POPROX}

The \textit{Platform for OPen Research and Online eXperimentation} (POPROX)\footnote{\url{https://poprox.ai}} is a community research infrastructure project funded by the U.S. National Science Foundation (NSF)\footnote{https://www.nsf.gov/awardsearch/showAward?AWD\_ID=2232551}. Our aim is to build a live recommendation platform that can host research studies examining real-world interactions between users and recommender systems, enabling the kind of user-centered research that is commonplace in industry settings but difficult for academic researchers. We chose (and the NSF supported) news recommendation as the initial domain for POPROX because it is a recommendation domain of social import. With the number of US citizens receiving their news on digital platforms greater than 50\% \citep{pew-news-platforms-2024}, concerns have rightly arisen about how the change from editorial curation of news to algorithmic curation impacts core democratic functions.

The POPROX platform has been live since January 2025, delivering personalized newsletters containing articles from the Associated Press (AP) to subscribers and supporting researchers investigating various aspects of recommendation. We chose a daily newsletter as the initial implementation of POPROX to be able to start studying real user responses quickly by leveraging the existing distribution and notification mechanisms of e-mail. It is also a relatively forgiving modality for recommendation generation, because researchers do not have to meet strict latency constraints in delivering recommendations and incorporating user feedback.

Since POPROX provides an existing pool of subscribers who are consented research subjects, researchers running experiments on the platform do not need to recruit experiment participants, secure their informed consent, or manage subscriptions. Although POPROX is a non-commercial research platform, maintaining a population of active subscribers exposes it to similar pressures as commercial products and industrial recommender systems. In particular, the ``market logic'' identified by \citet{mitova2023worlds} makes ``funnel thinking'' that considers and addresses issues of reach, engagement, conversion, and retention \citep{vandenbroucke2024ctr} relevant for the platform.

\section{News Recommendation Research}

News recommendation has a 30+ year history, stretching back to \textit{The Krakatoa Chronicle} \citep{krakatoa-chronicle}, an early web-based personalized newspaper. In that time, it has been the focus of long-running workshops like INRA\footnote{https://research.idi.ntnu.no/NewsTech/INRA/} and competitions such as the CLEF NewsREEL Challenge \citep{hopfgartner2015newsreel}, the RecSys Challenge \citep{kruse2024challenge}, and the MIND News Recommendation Competition\footnote{https://msnews.github.io/competition.html}. We highlight and summarize a few aspects of the literature most relevant to our work on POPROX below, and direct readers to a number of survey papers that cover this research area more fully \citep{karimi2018evalsurvey, li-survey-2019,feng-survey-2020, raza-survey-2022, meng-survey-2023, modeling-survey,iana-survey-2024, bauer-values-2024}.

\subsection{Datasets}

Available public news recommendation datasets in 2025 include MIND (English) \citep{mind-dataset-2020}, EB-NeRD (Danish) \citep{ebnerd-dataset-2024}, Adressa (Norwegian) \citep{adressa-dataset-2017}, plista (German) \citep{plista-dataset-2013}, NPR (Portuguese) \citep{npr-dataset-2023}, and IDEA (English) \citep{idea-dataset-2024}. Other datasets commonly used in the news recommendation literature are not currently available to researchers, either because they are no longer provided (Globo \citep{globo-dataset-2018}, Yahoo! Webscope) or because the data is proprietary (MSN News, Google News, Yahoo News).

Although the data contained by public news recommendation datasets varies somewhat, it typically includes user and article identifiers, article categories or topics, named entities mentioned by articles, and textual features such as article headlines, abstracts, and body text. Some datasets have been extended into multi-modal formats with the addition of images (e.g. IM-MIND \citep{im-mind-dataset}, VMIND \citep{vmind-dataset}).

\subsection{Modeling}

Recent work in news recommendation modeling displays two major trends (which are not mutually exclusive and are sometimes combined): (1) text-based models applying techniques from natural language processing (e.g. BERT embeddings) to article headlines, abstracts, and/or body text, and (2) knowledge-aware models incorporating knowledge graphs and/or non-text attributes (e.g. categories or named entities).
We refer readers to the modeling-oriented survey by \citet{modeling-survey} for detailed analysis of the model architectures and features used in news recommendation.

\subsection{Evaluation}

As is common throughout the field, most evaluation of news recommendation methods is performed offline using public or proprietary datasets, while online evaluation, A/B tests, and user studies are comparatively rare. Evaluation metrics are typically focused on predictive accuracy, with relatively few papers attempting to quantify ``beyond accuracy'' aspects like diversity, novelty, or serendipity. We refer readers to the evaluation-oriented survey by \citet{karimi2018evalsurvey} for a more detailed analysis of evaluation practices in the news recommendation literature. However, we highlight two noteworthy exceptions to these trends:

\begin{itemize}
\item CLEF NewsReel, a ``living'' evaluation lab for online and stream-based evaluation using plista's Open Recommendation Platform, where recommendations generated in response to live or replayed requests were required to adhere to realistic time constraints for real-world systems \citep{hopfgartner2015newsreel}.
\item Informfully, a news recommendation platform that includes a mobile app with which experimenters can conduct user studies on users they recruit and manage themselves \citep{heitz2024informfully}. Unlike POPROX, Informfully does not provide an associated pool of regular users; experimenters need to recruit their own research subjects.
\end{itemize}

\subsection{Values}

News recommendation presents a classic multistakeholder recommendation problem \citep{abdollahpouri2020multistakeholder}: we expect that recommendation platforms will have journalistic objectives distinct from the goals that users might have relative to personalized content \citep{moller2024designing,mitova2023worlds}. This need is quite explicit for news organizations that have licensing requirements which include support for a statutorily-defined public interest \citep{sorensen2019public,grun2023transparently,talleraas2020cultural}. Researchers have attempted to quantify and represent various journalistic objectives, especially news diversity, to be pursued in tandem with personalization \citep{heitz2022benefits,helberger2021democratic,vrijenhoek2022radio}. 

As a tool for experimental evaluation of recommender systems and especially because of its survey capabilities, POPROX provides the opportunity to explore the consequences of algorithmic choices for journalistic values and users' experience of current events. 

\section{Practical Challenges}

While the POPROX platform allows researchers to develop and deploy their own recommenders to support experiments, the system also needs to provide a satisfactory default news recommendation experience in the personalized newsletters it delivers to subscribers each day. This default experience serves two important purposes: supporting our efforts to recruit, engage, and retain a long-lived participant pool; and providing a solid base for researchers to use in building their own recommenders.

In this section, we outline some of the challenges we faced in building the default experience for POPROX and highlight where we were (and were not) able to rely on solutions from the research literature. We also describe our current solutions and their consequences for the platform.

\subsection{Training A Recommendation Model}

\paragraph{Key Issues}
In order to engage and retain users, the platform's default recommendation experience should present relevant news articles to POPROX subscribers. Relevance modeling is an important component of a recommendation system that does so, but since POPROX is a brand-new platform, we have not yet collected sufficient user behavior data to form a dataset that can be used for training or offline evaluation of recommendation models. 

Moreover, our system both collects and uses data types that are not present in public datasets. Beyond article text, our AP news feed contains a variety of metadata, including named entities and topics, but the tags provided with Associated Press articles are substantially different from what is present in MIND and other available datasets. Our system also allows users to express explicit topic preferences that are not present in any public dataset as far as we know. We hoped to train a recommendation model that takes advantage of these data types to better understand which articles are relevant for which readers.

\paragraph{Relevant Literature}
While headlines are commonly used as inputs to news recommendation models, categorical features like named entities and subject/topic categories are provided in many public datasets. While some models do use combined textual and categorical features, we found few options for models that used all of these input types in combination and were suitable for deployment in a live system.

Models that incorporate a wide range of input data beyond headlines tend to be graph-based models that learn embeddings or weights directly for user and/or item IDs, instead of using mechanisms to compose user and item representations from history and content features. Models with article ID features are viable for offline datasets with static content pools, but do not provide a viable recommendation strategy in a live system that must select from fresh items each day. Models with user ID features are potentially workable in the context of a live system, but would require online learning approaches or daily retraining/fine-tuning to keep user interest profiles up to date, all of which are beyond the current capabilities of the POPROX system.

\paragraph{Our Approach}
In end, we chose to train the NRMS model \citep{wu-etal-2019-nrms} on the MIND dataset, relying on transfer learning from MIND to AP data, which has worked well enough to get us started. NRMS encodes news articles by embedding the words in their headlines, contextualizing the word embeddings with self-attention, and condensing them into a single article embedding with additive attention. Users are encoded similarly based on their clicked news articles by compressing article embeddings into a single user embedding with similar attention mechanisms. Candidate articles are then scored via dot products between user and article embeddings in a fashion similar to matrix factorization.

\paragraph{Consequences}
The NRMS user and news encoders solely use article headlines, so we are unable to make use of much of the article data that we receive in our AP news feed. Various work-arounds that we considered, such as augmenting headline text with metadata in training, were not workable with the MIND dataset because its metadata is different from what AP provides. Our experience with the recommended content suggests that the model's single-embedding user representation may be biased toward some topics over others, and not well-suited to representing interest in multiple distinct and potentially non-overlapping topic areas. NRMS does not provide an obvious way to incorporate explicit user preference signals, so we have developed some workable-but-not-ideal approaches to providing such functionality in order to support user onboarding, which we describe below.

\subsection{Providing User Preference Controls}
\label{sec:usercontrol}

\paragraph{Key Issues}

New users sign up for POPROX and expect to get something reasonable in their news feed right away, which presents a risk of low engagement, poor retention, and high churn in the subscriber base if we fail to meet that expectation. In our initial design work, we quickly determined that ``something reasonable'' meant allowing users to declare interests across news topics: some users want to see sports news every day, some want to avoid it altogether. 
Since POPROX is intended to be a platform for experimenting with a wide range of models and interfaces (including experiments with the user onboarding process), we had to represent preferences in a way that is not overly tied to any one specific model or experience. We also needed to have a way to elicit these topic preferences efficiently and to allow users to modify them over time.

\paragraph{Relevant Literature}

Industrial recommender systems from both publishers and news aggregators do commonly provide user interface controls that enable readers to explicitly declare their interests, but we have been unable to identify news recommendation research papers that incorporate declared user interests (like those collected during an onboarding process) as preference signals. We hypothesize that this may be related to the prevalence of implicit feedback and unavailability of such explicit preference features in public datasets. We also did not find good empirical evidence for the relative effectiveness of different preference elicitation strategies, in news or other recommendation domains.

We did find strictly attribute-based recommenders, where users can give explicit feedback on attributes, and implicit feedback (item clicks) is decomposed into attribute weight adjustments based on the attribute values of the clicked items~\citep{knijnenburg2011each}, and there are examples of using query-specific attribute information for filtering recommendations \citep{loepp2015blended}. None of these use cases involve the creation of explicit standing preferences to be integrated into the recommendation process.

We also found reranking approaches for improving the calibration of recommendations relative to categories of user interest. For example, the greedy reranking algorithm in \citep{steck2018calibrated} uses a topic distribution of the user's ratings to produce recommendations that are distributed in genre similarly to the user's profile. This technique could be extended to a topic distribution derived from explicit preferences.

\paragraph{Our Approach}
To elicit topic preferences, the POPROX interface allows subscribers to select from a set of 14 high-level topics that align with the sections provided at the top of the AP News website.\footnote{https://apnews.com/} Subscribers indicate their interest in these topics using 5-point response scales ranging from ``Not at all interested'' to ``Very interested''. In the interest of providing users a measure of control over the articles they receive, we also allow subscribers to edit these preferences throughout their subscription period. 

In order to make user preferences influence recommended content, we added a separate topic-based ranking pipeline whose output is merged with a click-based ranking pipeline before selecting recommended articles. The topic-based pipeline treats textual definitions of AP topics as headlines and shoehorns them into the NRMS article embedding space with the model's news encoder, then applies the remainder of the model as usual to produce an estimated topic-based relevance score.

In designing the topical ranker pipeline, we have drawn inspiration from the use of negative feedback in the NRNF model \citep{nrnf-wu-2020} since it is difficult to represent positive and negative signals together in the same user embedding. We therefore use separate embeddings and scorers to estimate interest and disinterest, and the resulting interest and disinterest scores are then combined via subtraction.

\paragraph{Consequences}
With an initial implementation of topic preferences in place, we now face the challenge of ensuring that they work as expected. Our topical ranker pipeline seems to work acceptably well for users with narrow interests, but it is unclear how it could be extended. We would like to allow for more open-ended means of interest expression including named entities, locations, and others but we anticipate significant effort would be required to link those entities with descriptions or definitions (e.g. from an external knowledge base.) We are also not confident that a single-embedding user representation can adequately reflect a broader range of finer-grained interests.

Beyond the technical challenges, determining an appropriate blend of explicit preferences with implicit feedback is not straightforward. The recommender should honor user preferences to some extent and updating preferences should affect what recommendations are delivered in a noticeable way, but few offline accuracy metrics account for fidelity to explicit preferences. Furthermore, giving users more control over the news they receive may come at the cost of recommending informative content that, taken as a whole, adequately embodies journalistic values and fulfills the important roles of news providers beyond user engagement. For these reasons, we continue to rely in large part on ``taste-testing'' the recommendations but do not find this approach entirely satisfactory.

\subsection{Combining Curation With Personalization}

\paragraph{Key Issues}
News publishers and platforms take editorial stances not only in the news they cover and feature but also in the ways that news content is structured and presented for readers. Although recommendation technology expands the range of possibilities, the specific ways that personalization is deployed still reflects chosen resolutions of tensions between different logics and values, such as the tension between user engagement and the duty to inform. As a news recommendation research platform, POPROX is no different, and we are aware of the need to make informed, intentional decisions and to be explicit about the stances we take in designing the ways that articles are selected and displayed.

This presents significant practical challenges, since journalism and recommender systems have historically approached these issues in different ways. Newspapers and their digital equivalents have largely relied on the idea of sections, enabling curation across a range of diverse topics by providing different places to feature different kinds of news. Recommender systems research has heavily emphasized ranking a single recommendation list and mainly investigated ways to blend or balance multiple objectives when determining the order of items within. Finding appropriate ways to combine curation with personalization using these (and other) approaches remains an area of active investigation and exploration in both fields.

\paragraph{Relevant Literature}

We looked to several areas of the recommender systems and news recommendation literature for answers and approaches: multi-objective and multi-stakeholder approaches \citep{abdollahpouri2020multistakeholder}, grid/carousel interfaces \citep{rajOptimizingRankingGridlayout2024,rahdariCAREInfrastructureEvaluation2024, loepp2023carousel}, whole-page optimization \citep{dingWholePageOptimization2019}, values in news recommendation \citep{vrijenhoekRecommendersMissionAssessing2021a,strayBuildingHumanValues2024}, and algorithmic auditing \citep{lurieOpeningBlackBox2019}, among others. While each was informative and helpful in a general sense, we found few methods or results that were directly relevant to content-structuring approaches commonly taken by real-world news recommendation platforms and products, such as vertical sequencing of the ubiquitous ``Top News'' and ``For You'' sections (displaying curated and personalized content respectively) or the use of topical sections matching declared user interests.

\paragraph{Our Approach}
The current structure of POPROX newsletters contains a single ranked list of news articles selected for each user by our default recommender or by an experiment recommender (when an experiment is active on the platform). To the extent that we currently have an approach to combining editorial curation and recommender-based personalization, it is that we rely on the Associated Press to provide content that reflects their editorial standard and stances and apply a layer of personalization, resulting in a personalized selection of news from a curated content pool. We would like to move beyond this but still have many open questions about how to do so.
Our AP news feed provides headline packages featuring the top ten stories for each of a range of topics (e.g. US News, Sports, Entertainment) but does not provide such a list of the overall top stories of the day. As a result, we are not able to fully rely on their editorial curation to determine what to feature as top news in POPROX newsletters, and would need to apply some form of (personalized or non-personalized) algorithmic selection.

\paragraph{Consequences}
On one hand, presenting only personalized content as we currently do could result in some degree of ``filter bubble'' issues \citep{vanalstyneGlobalVillageCyberbalkans2005, flaxman2016filter,michielsWhatAreFilter2022,pariser2011filter}. We expect these may be mitigated somewhat by recommending news articles from a fairly limited pool, since it is difficult for subscribers to delve too deeply into any single topic when the number of available articles per topic is low.

On the other hand, a potential future structure of the newsletter that incorporates multiple sections would become quite difficult to evaluate using standard accuracy-oriented offline evaluation techniques designed for single top-$K$ lists. While the POPROX platform does provide a suite of tools for online and offline evaluation that includes some alternatives, we are hesitant to make changes to the newsletter that would deprive experimenters using the platform of familiar and useful tools without providing a workable substitute, which would present its own research, development, and validation challenges.

\subsection{Assessing User Experiences And Satisfaction}
\paragraph{Key Issues}
Online and offline evaluation methods based on content properties and behavioral tracking are necessary but not sufficient for understanding how recommendations are experienced by their recipients. User surveys therefore form an essential part of our platform, and provide several benefits:

\begin{itemize}
\item Because many newsletter consumers simply read the headlines without clicking through to the linked articles, users' satisfaction with the recommended news articles may not always be apparent from their interaction behavior. Surveys provide an additional signal of these ``passive'' users' satisfaction with their recommendations.
\item Users may be interested in an article for several reasons, and a certain algorithm may be particularly good at fulfilling one specific type of user need (e.g. feel-good articles) or at catering to a balanced set of needs/interests. A survey can provide more contextual granularity to users' evaluations that can then be triangulated with their behavior~\cite{knijnenburg_explaining_2012,knijnenburg_evaluating_2015}.
\item Surveys can cover constructs related to the long-term effects of a recommender system, such as its ability to help users explore, understand, and develop their interests in a variety of news topics~\cite{knijnenburg_recommender_2016}. They can also include key user demographics and characteristics that can be used to evaluate the fairness of proposed recommendation algorithms and other interventions.
\end{itemize}

An important experimental advantage (but methodological challenge) of POPROX is the perpetual and longitudinal nature of our studies and therefore of our evaluations. Whereas most user-centered research studies in recommender systems involve short-lived interactions with a system, POPROX applies interventions to a user base that has ongoing interactions with our platform. This increases the realism and ecological validity of implemented studies and allows researchers to track the effects of interventions over time, but also complicates holistic evaluation because surveys must be administered perpetually (before, during, and after a study).

\paragraph{Relevant Literature}

To our knowledge, little to no published research has considered the longitudinal dimension of user experience in recommender systems, let alone the methodological question of how much time it takes for an intervention to ``take hold'' and how long an effect may linger post-intervention.
Furthermore, while there exists a vast body of research on increasing survey participation in human subjects research~\cite{groves_understanding_1992}, our platform much more closely resembles commercial settings in this regard, and much of the research in those settings is proprietary and unpublished.

\paragraph{Our Approach}

The POPROX platform sends users a short weekly survey, which rotates on a 5-week cycle through constructs related to users' perceptions, experience, and need fulfillment (which is at some occasions measured across the newsletters of the past week, while at other times we ask users to evaluate these constructs for a specific newsletter).
On the fifth week of the cycle, the survey system rotates through a series of personal characteristics and demographics, which can be used to select/stratify a user sample (e.g. balance gender in a study, or target users of a certain age group), or to contextualize evaluations (e.g. evaluate the effect of an algorithmic intervention across users with different personalities).
To reduce survey length, we measure each construct with a single item taken from pre-validated scales.

\paragraph{Consequences}
While we have been able to rely on Qualtrics for core survey functionality, we had to devise our own solutions for creating a rotating schedule of periodic surveys involving nested cycles. Due to the voluntary, unpaid nature of users' participation in the POPROX platform, we have sought to balance user time and effort with our goal of collecting robust data on multiple user experience constructs. Despite our efforts to minimize their length, our initial deployment has shown very low engagement with weekly surveys. We are investigating whether we need to incentivize survey completion in order to improve the ``conversion rate'' of active readers into survey respondents.

\section{Lessons for Future Research}

Our work building POPROX and experience attempting to locate and apply research findings to support this effort leads us to several lessons for news recommendation as well as for the broader RecSys community.
We expect many of these lessons may be unsurprising to people already working on production news recommenders, but we find them often missing from the literature, and therefore valuable to explicitly articulate for the community and for new researchers and practitioners looking to enter the field.

\paragraph{Treat using the available data as a first-order concern.}

Our challenges highlight several interrelated issues with the ways that data is used and discussed in research:

\begin{itemize}
\item Limiting modeling or evaluation to a subset of the attributes (or instances) in a dataset prevents the dataset's contents and capabilities from serving as a forcing function for developing adaptable models and systems.
Accommodating data types beyond least-common-denominator attributes like clicks requires more flexibility than many of the models found in the literature provide.
\item Rich item-level data is a natural substrate for facilitating user preferences and feedback, editorial control, business rules, and other functionality that provides human influence on system behavior.
When little research is available on the use of a particular type of data, it is difficult to construct or extend recommender systems to provide user and stakeholder controls built on that data.
\item Many papers do not provide clear and thorough details on data preparation for modeling and evaluation, as others have noted in the context of evaluation rigor \citep{sunAreWeEvaluating2020, tanCriticalStudyMovieLens2023}. Thoroughly documenting data preparation decisions, including splitting, missing value imputation, feature engineering, etc. subjects data decisions to peer-review, allowing for feedback and community vetting of data practices. It also aids reproducibility and provides readers with worked examples of effectively using data, helping students and new researchers learn.
\end{itemize}

Given the importance of data to recommendation and evaluation, promoting more research on data seems promising as a path to impactful improvements in recommender systems that transcend domains and model families.
There are some examples in the literature, such as work on feature engineering \citep{schifferer2020challenge, verdonck2024features}, data minimization \citep{biega2020operationalizing,sonboli2024tradeoff} and the study of fairness research practicalities by \citet{daniilChallengesStudyingBias2025}.

\paragraph{Model affordances matter.} Real-world recommender systems must be developed with a degree of ``mechanical sympathy'' for how the models in use work, their strengths and weaknesses, and the interactions between those models and other system components. This is more straightforward with models that provide clear indications of how they can be used or extended to support common system functionality, and more difficult when those indications are lacking. In this regard, both domain-specific and domain-independent recommendation models often leave something to be desired.

On one hand, a domain-specific model may focus the proposed architecture solely on the features commonly available in public datasets from that domain. For example, many news recommendation models use only textual input features and require architectural modifications in order to accept categorical inputs. Others incorporate one specific categorical input (like topics) in a way that is difficult to extend to other categorical features. On the other hand, general purpose click-through rate prediction models accept a wide range of continuous and categorical features but provide few indications about how to construct appropriate input features to support desired system functionality (in general or in specific domains) even when that is possible and well-supported in practice.

We view these issues, at least in part, as a reflection of the field's emphasis on predictive accuracy over examining how models fit into the broader context of real-world systems, where models may either help or hinder practitioners in building systems that embody the requirements, values, and user experiences they aim to realize. We believe this represents an opportunity to investigate and improve the ways that recommendation models support common system design patterns and signal that support to recommender system developers.

\paragraph{Recommendation is more than modeling.} Research and evaluation findings often focus on the scores produced by a model, or on rankings derived from those scores. However, in practice the quality of delivered recommendations depends on much more than accuracy, or even non-accuracy properties, of models and their outputs. A personalized news application may:
\begin{itemize}
\item Inquire about users' interests and disinterests to collect explicit preference signals
\item Select candidate items from multiple sources centered on different news publishers, subjects, or geographical locations
\item Exclude items from consideration based on user- or platform-defined criteria (e.g. ``Hide all stories from [this source]'')
\item Estimate the likelihood of types of user engagement such as reading, saving, and sharing
\item Recommend content across several modalities including articles, podcasts, and videos
\item Satisfy list construction and ordering constraints imposed by product requirements (e.g. putting top news first, including certain percentages of local/national/international news)
\item Balance multiple objectives for different stakeholders including readers, editors, journalists, and advertisers
\end{itemize}

Implementing this range of functionality requires a significant amount of software that is not frequently described in the literature or supported by recommender system frameworks or toolkits.
To the best of our knowledge, there are only two academic or industrial frameworks (NVIDIA Merlin \citep{higley2022merlin} and recent versions of LensKit \citep{lenskit-framework}) that provide tools for building more complex multi-stage pipelines that combine models with other components, and neither offers specific support for news recommendation. As part of the POPROX platform, we aim to provide such a toolkit\footnote{https://github.com/CCRI-POPROX/poprox-recommender} for news recommendation researchers in order to support their efforts to build and evaluate news recommenders that deliver recommendations to users in the context of a live system.

\paragraph{Live systems present opportunities to apply a spectrum of evaluation methods.} There are many well-known and frequently used evaluation approaches, including: accuracy and ``beyond accuracy'' offline evaluation metrics; online behavioral metrics and A/B testing; and user studies and surveys. We see opportunities to bridge the gaps between these methods by expanding both the system components included in evaluations and the ways that those components are evaluated (independently and together). We highlight three that are relevant for our system and that we believe are applicable in many contexts:

\begin{itemize}
\item \textbf{Correctness testing:} The idea of evaluating whether a model or recommender meets particular functional requirements beyond statistical evaluation is relatively new to the field, but \citet{michielsFrameworkToolkitTesting2023} proposed the idea of recommender system test suites for specific behavior, which could be extended with domain- or application-specific behavioral evaluations and acceptance tests.
\item \textbf{End-to-end offline evaluation:} Live systems with personalization features often use multi-stage recommenders, including retrieval and ranking models in conjunction with filters, re-rankers, and business logic. With appropriate system construction, these additional components can be included in offline evaluations to measure the accuracy and ``beyond accuracy'' impacts of design choices within and between these system components.
\item \textbf{Distributional evaluation:} In real-world contexts, we are often concerned not only with aggregate performance and quality metrics but also with identifying which users and items are well-served or under-served in order to ensure that recommendations meet a minimum quality bar. For example, we have noticed biases in our system toward certain topics and away from others that we believe may originate from imbalances between topical categories in training data. Assessing such biases and disparities in utility across different categories and stakeholders calls for examining not only point estimates but also distributions, as suggested by \citet{ekstrand-distributional-eval-2024}.
\end{itemize}

\section{Conclusion}

In this case study, we have described our experiences creating POPROX and some of the challenges we faced in doing so. Although news recommendation has been an active topic of research for several decades, we did not receive as much help from the research literature as we hoped. Part of this gap was expected, as we knew that news recommendation has unique characteristics among recommendation domains and that creating a long-running live recommender system as research infrastructure was not something that many research groups had attempted. Part of the gap was surprising though, since some features that are commonplace in commercial news recommendation products and platforms remain under-researched and have proved difficult to implement with published techniques.

We believe there are multiple reasons why these challenges exist. One is certainly the focus on the ``horse race'' of chasing an evaluation benchmark on a narrowly defined task. This kind of focus can lead to methods and models lacking functional characteristics that would make them suitable for deployment in real systems. We have also often encountered a push in recommender systems research (through reviewing, community expectations, etc.) to show that findings are generalizable: that they apply across multiple data sets, domains, and/or applications. Effectively building recommender systems that serve real users, however, requires deep and specific engagement not just with a domain in general, but with the particular characteristics of specific datasets, applications, and user communities.

While we recognize and have encountered significant systemic challenges and barriers, we nonetheless encourage more researchers to start, or get involved in, longer-term research infrastructure projects that build and operate live recommender systems. Engaging in a multi-year, interdisciplinary recommender systems effort provides a catalyst for integrating approaches from many areas and fields, which also serves to highlight what is missing. In the context of a long-lived system, different categories of issues surface that would not arise in a standalone research project over the course of a semester or a year.

In aggregate, these issues and gaps offer a different perspective on the extent to which RecSys research is making cumulative progress: while the field is making considerable advances on a number of important but narrowly defined recommendation problems, these advances do not yet ``add up'' to the knowledge base that is needed to build a real-world recommender system. By encouraging more researchers to engage in complex real-world projects, we hope to spur the RecSys community to become more aware of and attentive to research that fills the gaps that stand between the current state of the field and greater real-world impact.

\begin{acknowledgments}
This work is based on research supported by the National Science Foundation under Grant Nos. IIS 22-32551, 22-32552, 22-32555, and 24-09199. We are grateful to the rest of the POPROX team for their ongoing collaboration in this effort and many discussions that have informed this paper.
\end{acknowledgments}

%% The declaration on generative AI comes in effect
%% https://ceur-ws.org/GenAI/Policy.html
\section*{Declaration on Generative AI}
  The author(s) have not employed any Generative AI tools.

%%
%% Define the bibliography file to be used
\bibliography{beyond}

\end{document}